\documentclass[pre,showpacs,preprint]{revtex4}
\usepackage{graphicx}
\usepackage{amsfonts}
\begin{document}

\title{Numerical validation of the complex Swift-Hohenberg equation for lasers}

\author{Juan Pedrosa$^1$, Miguel Hoyuelos$^{1,2}$ and Carlos Martel$^3$, }

\affiliation{
$^1$Departamento de F\'{\i}sica, Facultad de Ciencias Exactas y Naturales,
Universidad Nacional de Mar del Plata, Funes 3350, 7600 Mar del
Plata, Argentina \\
$^2$ Consejo Nacional de Investigaciones Cient\'{\i}ficas y T\'{e}cnicas,
CONICET, Argentina\\
$^3$Departamento de Fundamentos Matem\'{a}ticos, ETSI
Aeron\'{a}uticos, Universidad Polit\'{e}cnica de Madrid, 28040 Madrid, Spain}

\begin{abstract}
Order parameter equations, such as the complex Swift-Hohenberg (CSH)
equation, offer a simplified and universal description that hold
close to an instability threshold. The universality of the
description refers to the fact that the same kind of instability
produces the same order parameter equation.  In the case of lasers,
the instability usually corresponds to the emitting threshold, and
the CSH equation can be obtained from the Maxwell-Bloch (MB)
equations for a class C laser with small detuning. In this paper we
numerically check the validity of the CSH equation as an
approximation of the MB equations, taking into account that its terms
are of different asymptotic order, and that, despite of having been
systematically overlooked in the literature, this fact is essential
in order to correctly capture the weakly nonlinear dynamics of the
MB. The approximate distance to threshold range for which the CSH
equation holds is also estimated.
\end{abstract}


\maketitle

\section{Introduction}

The complex Swift-Hohenberg (CSH) equation is an order parameter
equation that provides a reduced description of a variety of systems
\cite{crosshohenberg93}, such as Rayleigh-B\'{e}nard convection
\cite{swifthohenberg77}, optical parametric oscillators
\cite{LonghiGeraci96,SantagiustinaHernandezSanMiguelScroggieOppo02,HewittKutzt05},
Couette flow \cite{Manneville04}, nematic liquid crystal
\cite{Buka04}, magnetoconvection \cite{Cox04}, propagating flame
front \cite{Golovin03} and photorefractive oscillator
\cite{Staliunas95,StaliunasSlekysWeiss97} among others. In the case
of lasers operating near peak gain (small detuning), a derivation of
the CSH equation for class A and C lasers was obtained in
\cite{LegaMoloneyNewell94,LegaMoloneyNewell95}, starting from the
semiclassical Maxwell Bloch (MB) equations
\cite{Risken68,Narducci86,lugiato88,Coullet89,Jakobsen92}, that
provide a general description of transverse patterns in two levels,
wide aperture and single longitudinal mode lasers.  For class B
lasers, such as CO$_2$ and semiconductor laser, CSH equations have
been obtained in \cite{Barsella99} and \cite{Mercier02} respectively.
(For an explanation of the classification of lasers, see Ref.\
\cite{Narducci88}.)  Experimental observations of patterns in wide
aperture lasers were reported in, for example,
\cite{Dangoisse92,Heckenberg92,DAngelo92,Coates94,Lippi99}.  (For
reviews on pattern formation in nonlinear optical systems, see
\cite{ArecchiBoccalettiRamazza99,Lugiato99,StaliunasSanchez03}.)

The deduction of a generic order parameter equation greatly
simplifies the theoretical description of the system.  But it is
important to note the limitation of this kind of model equations, as
Cross and Hohenberg state in their review \cite[p.\
874]{crosshohenberg93}: ``It is true that many properties of
nonequilibrium systems are encountered in these equations, and indeed
many hard problems (...) may profitably be addressed in the simple
framework provided by these equations. However, it is only as a
perturbative expansion valid in a small region near threshold that
they provide a quantitative description of real experimental systems,
and results may be even qualitatively misleading if applied far from
threshold.''

We will focus our attention in a class C laser near peak gain that is
assumed to be well described by the corresponding Maxwell Bloch
equations.  Using the assumptions that the amplitudes of the physical
fields are small and depend slowly on time and on the transversal
spatial scales, a CSH equation is derived form the Maxwell Bloch
equations. {\bf The resulting CSH equation is that derived by Lega
et. al. in \cite{LegaMoloneyNewell94,LegaMoloneyNewell95}. Although
not explicitly stated in this original formulation, the equation
includes terms of different asymptotic order, in contrast to the
Ginzburg Landau equation that is obtained for negative detuning
\cite{NewellMoloney92}}.  This asymptotic nonuniformity is {\bf
systematically} obviated in the literature and was only recently
addressed \cite{martelhoyuelos06}. It is the manifestation of the
fact that dispersion and diffusion have {\bf necessarily different
asymptotic order, and it affects the slow scales that the system
develops near threshold.}

The argument of a qualitative only scope of the model equation is
usually invoked to justify the application of the CSH equation far
from threshold.  The qualitative correctness is difficult to be
theoretically established, but, on the other hand, the capacity to
produce quantitative predictions can be determined from the numerical
integration of both, the original system of Maxwell Bloch equations
and the CSH equation. The results of the comparison between the
numerical simulations of the CSH and the MB equations is what we
present in the subsequent sections of this paper.
  We can obtain from the simulations the relative error that
introduces the approximation and estimate how far from threshold we
can increase the pump while keeping a small relative error. Also,
this numerical comparison between the CSH and the MB equations
provides a confirmation of the main result presented in
\cite{martelhoyuelos06}: {\bf that the CSH equation for the
description of the weakly nonlinear dynamics of the system near
threshold (derived in \cite{LegaMoloneyNewell94,LegaMoloneyNewell95})
necessarily contains terms of different asymptotic order.}

\section{The complex Swift-Hohenberg equation and its numerical phase diagram}
\label{csh}

The Maxwell-Bloch equations for a two-level single longitudinal mode
laser with flat mirrors are
\begin{eqnarray}
 & \frac{\partial E}{\partial t}=ia\nabla^{2}E-\sigma E+\sigma P,\label{MBE}\\
 & \frac{\partial P}{\partial t}=-(1+i\Omega)P+(r-N)E,\label{MBP}\\
 & \frac{\partial N}{\partial
 t}=-bN+\frac{1}{2}(\overline{E}P+E\overline{P}),
 \label{MBN}
\end{eqnarray}
where $E(x,y,t)$ and $P(x,y,t)$ represent the complex electric and
polarization fields, and $N(x,y,t)$ is the real valued field of the
population inversion (the same nondimensional formulation as in Ref.\
\cite{NewellMoloney92} is used).  Parameter $a>0$ is the strength of
the diffraction (that we set to 1 by scaling the space variables),
$\sigma>0$ is the cavity losses, $\Omega$ is the cavity detuning (the
difference between atomic and resonance frequencies), $r$ is the
pumping parameter, $b>0$ is the decay rate of the population
inversion, $\nabla^{2}=\partial^{2}/\partial
x^{2}+\partial^{2}/\partial y^{2}$ is the laplacian operator in the
plane transverse to light propagation, and the bar stands for the
complex conjugate.  We will consider, as a specific case, a class C
laser, for which $\sigma \sim 1$ and $b \sim 1$.

The corresponding CSH equation was obtained in
\cite{LegaMoloneyNewell94,LegaMoloneyNewell95}, and a simpler
derivation method,
in which no a priori relative scaling of the variables is assumed,
was introduced in \cite{martelhoyuelos06}.  A linear stability
analysis of the Maxwell Bloch equations shows that the lasing
instability takes place at a critical value of the pump $r_c = 1$.
The assumptions of small detuning and small distance to threshold,
$$ |\Omega| \ll 1, |r-1| \ll 1, $$
are expressed through a small parameter $0< \varepsilon \ll 1$:
\begin{eqnarray}
\Omega &=& \frac{(\sigma+1)^{2}}{\sigma}\omega\varepsilon, \nonumber
\\ r-1 &=& \frac{(\sigma + 1)^2}{\sigma^2}(\omega^2 + \alpha)
\varepsilon^2,
\end{eqnarray}
were $\alpha$ and $\omega$ are order 1 parameters that represent the
scaled pump and detuning respectively.

The resulting CSH equation is of the from:
\begin{equation}
\phi_{\tilde{t}}=\alpha\phi+\textrm{i}\nabla^{2}\phi-\phi|\phi|^{2}-
2\varepsilon\omega\nabla^{2}\phi-\varepsilon^{2}\nabla^{4}\phi,
\label{CSHscaled}
\end{equation}
where time and space were scaled as
$\tilde{t}=\frac{(\sigma+1)}{\sigma}t\varepsilon^{2}$ and
$(\tilde{x},\tilde{y})=\frac{(\sigma+1)}{\sqrt{\sigma}}(x,y)\varepsilon$.

{\bf This CSH equation is exactly the same as  that obtained by Lega
et. al. in \cite{LegaMoloneyNewell94,LegaMoloneyNewell95}, but with
the variables rescaled to show that it has terms of different
asymptotic order and that it is not possible to remove the small
parameter $\varepsilon$ from the equation. This asymptotic
nonuniformity comes from the simple fact that dispersion involves
second order spatial derivatives while double diffusion has fourth
order ones and thus, in the long wave approximation where higher
derivatives correspond to smaller terms, these two terms have
necessarily different asymptotic order. This crucial fact is
precisely what forced Lega et. al.
\cite{LegaMoloneyNewell94,LegaMoloneyNewell95} to derive the CSH
expanding first up to two orders (the first one included dispersion
and the next the double-diffusion) and then collapsing back the
expansion to get the CSH equation. But, despite of the wide use of
the CSH equation, the asymptotic nonuniformity is never mentioned in
the literature, and it was only recently analyzed in
\cite{martelhoyuelos06} where it was shown that it gives rise to two
characteristic slow scales: one associated with dispersion
$\delta_{\mathrm{disp}}$ and a second one associated with diffusion
$\delta_{\mathrm{diff}}$. Using the scaling indicated above,
$\delta_{\mathrm{disp}}\sim 1$ and $\delta_{\mathrm{diff}}
\sim\sqrt{\varepsilon}\ll 1$, but in the original scaling of the
Maxwell Bloch equation $\delta_{\mathrm{disp}}\sim 1/\varepsilon \gg
1$ and $\delta_{\mathrm{diff}} \sim1/\sqrt{\varepsilon}\gg 1$, so
both are long spatial scales.}

The CSH equation above has to be considered in the close-to-threshold
limit of $\varepsilon\rightarrow0$, and the relation between the
Maxwell Bloch and CSH solutions can be
written as
\begin{equation}
\left[\begin{array}{c}
E(x,y,t)\\
P(x,y,t)\\
N(x,y,t)\end{array}\right]
 = \left[\begin{array}{c} 1\\1\\0 \end{array}\right]
 \sqrt{b}\frac{(\sigma+1)}{\sigma} \textrm{e}^{-i\sigma\omega \tilde{t}/\varepsilon}
 \phi(\tilde{x},\tilde{y},\tilde{t}) \,
\varepsilon + O(\varepsilon^2), \label{EPNexpansion}
\end{equation}

\begin{figure}[t]
\begin{center}
\includegraphics[width=0.9\columnwidth,keepaspectratio]{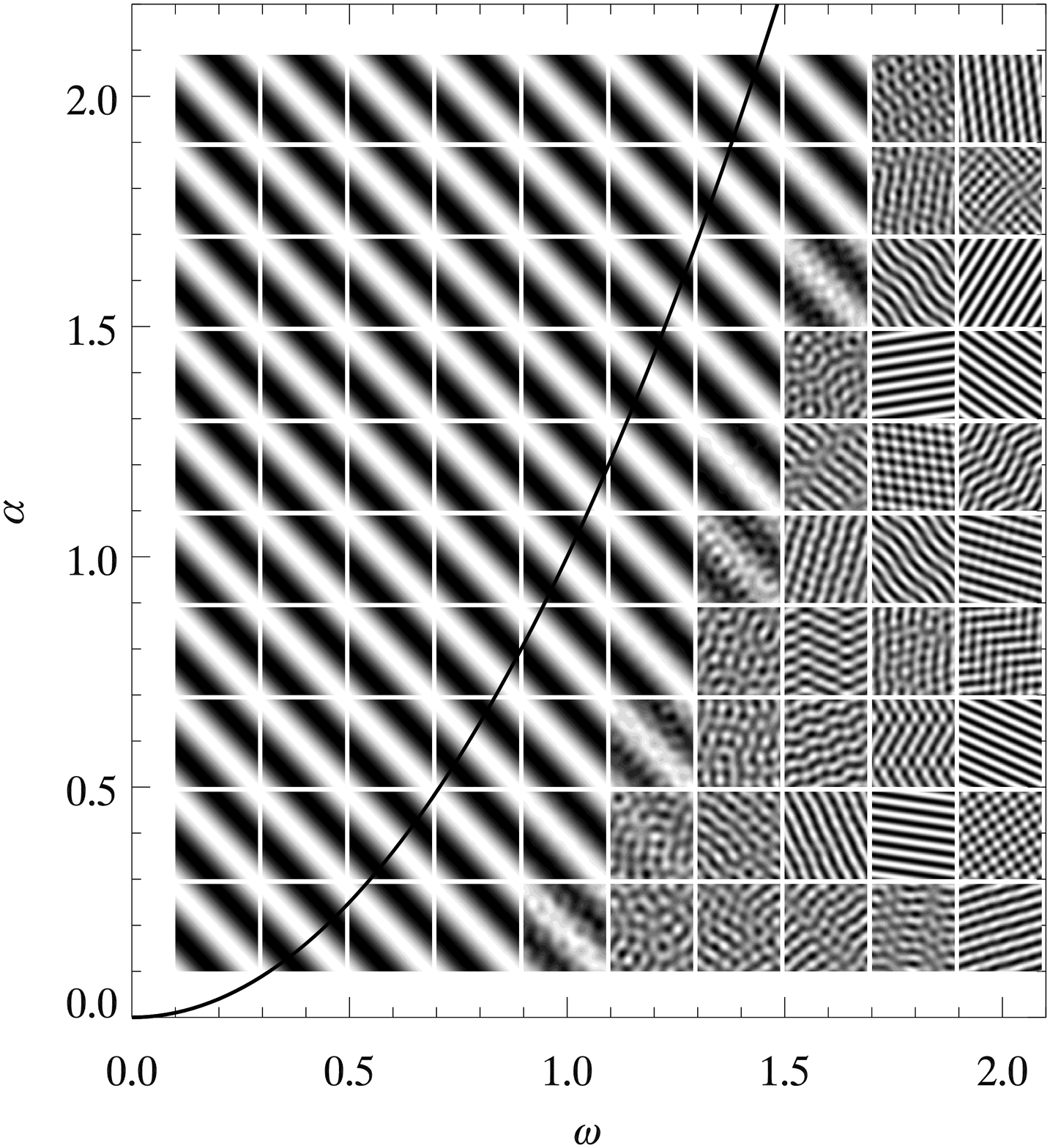}
\end{center}
\caption{\label{apos} Real part of $\phi$, in a 2D system of size
3$\times$3 with periodic boundary conditions, for different points in
the region of parameter space $\alpha>0$ and $\omega>0$. Each square
represents an individual numerical integration of Eq.\
(\ref{CSHscaled}) with initial condition given by a traveling wave,
with $\mathbf{k}_{\rm{TW}}=(1,1)\,2\pi/3$, plus noise of amplitude
0.02. The curve $\alpha = \omega^2$ is the stability limit of this
kind of solution.  The final time is $\tilde{t}=10$ and $\varepsilon
= 0.0083$. The gray scale limit values are: black $\simeq -2.5$ and
white $\simeq 2.5$}
\end{figure}

Traveling wave solutions of the form
$\phi_{\rm{TW}}=\sqrt{\alpha}\:\textrm{e}^{\textrm{i}\mathbf{k}_{\rm{TW}}
\cdot\tilde{\mathbf{x}}- \textrm{i}k_{\rm{TW}}^{2}\tilde{t}}$, with
$k_{\rm{TW}}=|\mathbf{k}_{\rm{TW}}|\sim1$ are approximate solutions
of the CSH equation up to $O(\varepsilon)$ corrections (this family
of TW is just the result of making the limit
     $\varepsilon\rightarrow 0$ and $k\sim 1$ in the well known expression of the
   exact TW family, see
   \cite{oppoalessandrofirth91,Jakobsen92,Jakobsenlegafeng94,
         LegaMoloneyNewell94,LegaMoloneyNewell95}).
  This solution exists only for $\alpha > 0$ and a linear stability analysis
shows that  it becomes unstable outside the region defined by $\omega
> 0$ and $\alpha > \omega^2$ with a critical wavenumber
$k_c=\sqrt{\omega/\varepsilon} \gg 1$ \cite{martelhoyuelos06}, which
corresponds to a perturbation with small {\bf diffusive} length
scale.  The phase diagram in the parameter space $\alpha$-$\omega$
represented in Fig.\ \ref{apos} was numerically reproduced starting
with an initial condition with $\mathbf{k}_{\rm{TW}}=(1,1)\,2\pi/3$
plus noise of amplitude 0.02. The system has {been integrated using}
periodic boundary conditions in a {square box of length 3}.
 The mesh of Fig.\ \ref{apos} represents the final
states for the corresponding values of $\alpha$ and $\omega$, for
$\tilde{t}=10$ and $\varepsilon = 0.0083$ (each square in the mesh is
the result of an individual numerical integration). To the right of
the parabola $\alpha = \omega^2$ the traveling wave solution becomes
unstable and gives rise to another structure with smaller wavelength
associated with the diffusive terms in Eq.\ (\ref{CSHscaled}). {
Some squares of the mesh still show the long wavelength solution to
the right of the parabola, where it should be unstable. The reason is
that, close to the stability limit, the unstable modes require a time
greater than $\tilde{t}=10$ to grow.}

\begin{figure}[t]
\begin{center}
\includegraphics[width=0.9\columnwidth,keepaspectratio]{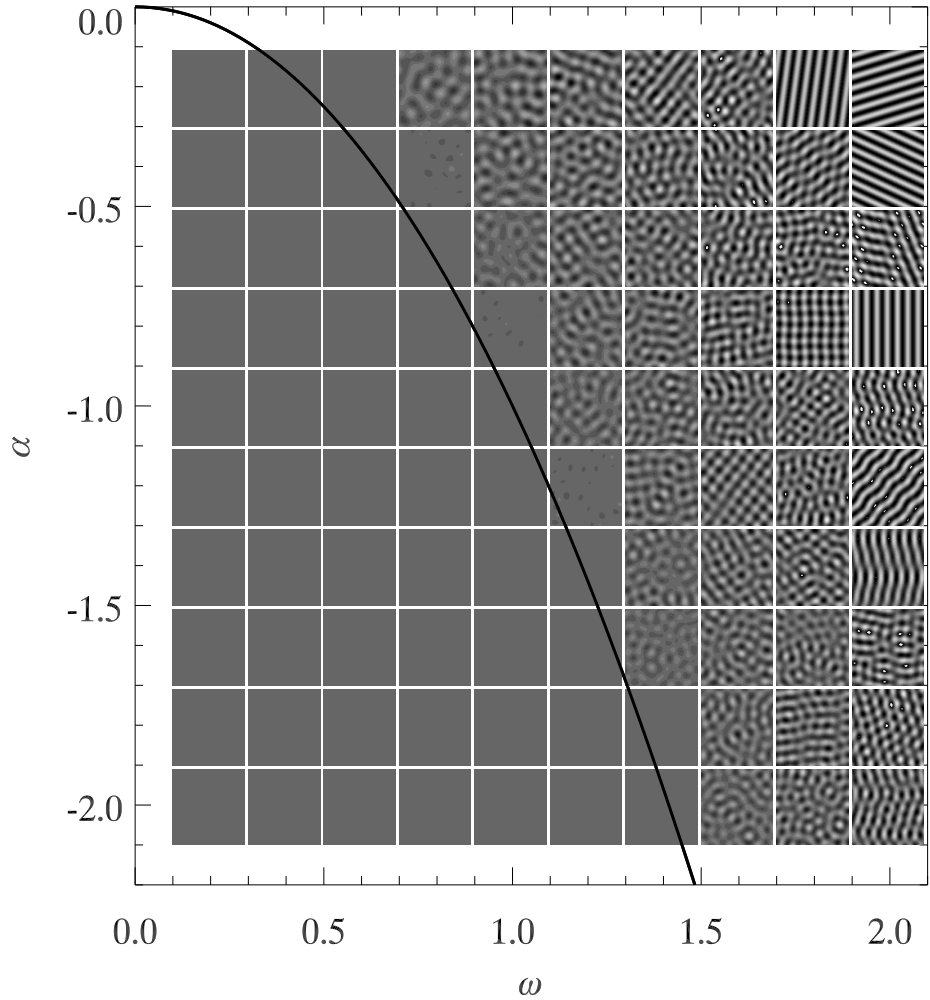}
\end{center}
\caption{\label{aneg} Real part of $\phi$ for different points in the
region of parameter space $\alpha<0$ and $\omega>0$. Each square
represents an individual numerical integration of Eq.\
(\ref{CSHscaled}) with initial condition given by the nonlasing
solution, $\phi=0$, plus Gaussian noise of amplitude 0.2. The curve
$\alpha = -\omega^2$ is the stability limit of the zero solution. The
final time is $\tilde{t}=10$ and $\varepsilon = 0.0083$. The gray
scale values are: black $= -2.5$, gray $=0$, and white $= 2.5$}
\end{figure}

The nonlasing solution, $\phi = 0$, is linearly unstable for $\alpha
> 0$ if $\omega < 0$, and for $\alpha > -\omega^2$ if $\omega >
0$ (exhibiting again a large diffusive critical wavenumber $k_c
=\sqrt{\omega/\varepsilon} \gg 1$) \cite{martelhoyuelos06}. The
numerical phase diagram of Fig.\ \ref{aneg} confirms again the
theoretical stability predictions and shows the appearance of a
structure with wavenumber $k \sim 1/\sqrt{\varepsilon} \gg 1$ to the
right of the stability limit given by the parabola $\alpha >
-\omega^2$. The initial condition is Gaussian noise with amplitude
0.2, the final time is $\tilde{t}=10$ and $\varepsilon = 0.0083$.

In the next section we will analyze the dynamics in two
representative points of the phase diagram: $\alpha = 0.75$, $\omega
= 0.5$ (pattern with dispersive scale $\delta_{\mathrm{disp}} \sim
1$), and $\alpha = 0.5$, $\omega = 2$ (pattern with diffusive scale
$\delta_{\mathrm{diff}} \sim \sqrt{\varepsilon}$).

\section{Numerical validation}

We numerically check the accuracy of the CSH equation
(\ref{CSHscaled}) as a reduced dynamics of the Maxwell Bloch
equations (\ref{MBE})-(\ref{MBN}).  The difference between both
descriptions is computed as
\begin{equation}
d = \left|\left| \left[\begin{array}{c}
E(x,y,t)\\
P(x,y,t)\\
N(x,y,t)\end{array}\right] - \left[\begin{array}{c} 1\\1\\0
\end{array}\right] \sqrt{b}
\frac{(\sigma+1)}{\sigma} \textrm{e}^{-i\sigma\omega
\tilde{t}/\varepsilon} \phi(\tilde{x},\tilde{y},\tilde{t})\,
\varepsilon \right|\right|.
\end{equation}
Symbols $||\cdot||$ denote the Euclidian norm on $\mathbb{C}^{3N}$
divided by $\sqrt{N}$, where $N$ is the number of points of the
discretized system. So, $d$ is an average absolute error, that,
according to the weakly nonlinear procedure applied to the MB
equations to derive the CSH equation, has to behave as $d \sim
\varepsilon^2$; see  Eq.\ (\ref{EPNexpansion}).

There is a {severe}  numerical difficulty in integrating Eq.\
(\ref{CSHscaled}) due to the presence of two {spatial} scales,
$\delta_{\mathrm{disp}} \sim 1$ and $\delta_{\mathrm{diff}} \sim
\sqrt{\varepsilon}$, which are very different in the relevant limit
$\varepsilon\rightarrow0$ and should be simultaneously well resolved.
We consider a one dimensional system, with periodic boundary
conditions, in order to reduce the size of the computations and be
able to use a greater system length than would be possible in higher
dimensions.  We let the system evolve until the difference $d$
reaches a stationary value $d_s$.  { For the parameters used, a
time $\tilde{t} \sim 10$ is enough to reach a stationary state, and
the corresponding} maximum {integration} time for the MB equations is
$t \sim 10/\varepsilon^2$. Therefore, to check the asymptotic
theoretical behaviour for $\varepsilon \rightarrow 0$, we need a
large number of Fourier modes and long time. The CSH and MB equations
are integrated in Fourier space using a fourth-order Runge-Kutta
scheme, with 1024 Fourier modes, $\sigma=b=1$, time step $dt = 0.01$
($d\tilde{t} = 2\,dt\,\varepsilon^2$), space step $d\tilde{x} =
1/1024$ [$dx = d\tilde{x}/(2\varepsilon)$], and $\varepsilon$ in the
range between 0.0011 and 0.025.

\begin{figure}[t]
\begin{center}
\includegraphics[width=0.8\columnwidth,keepaspectratio]{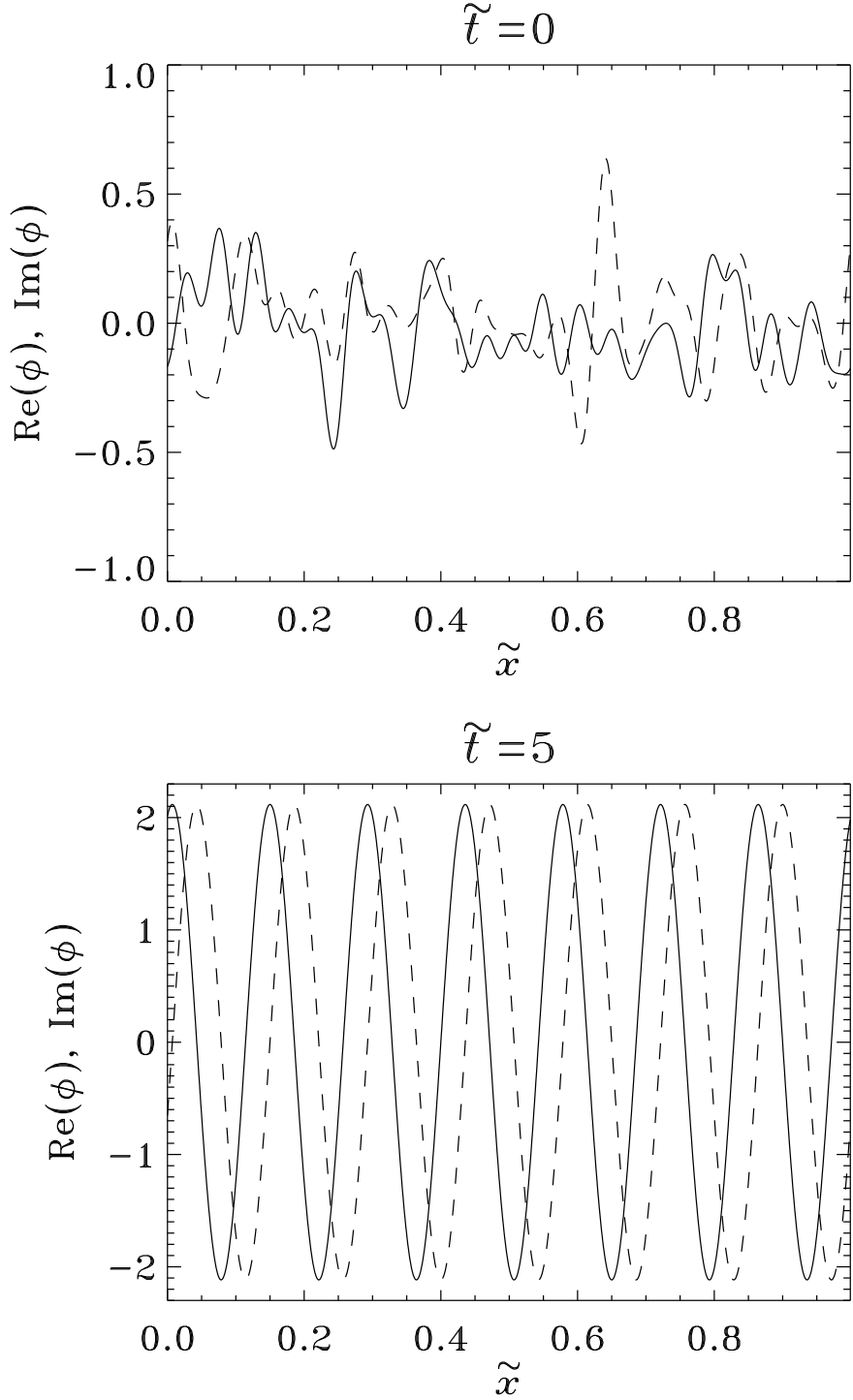}
\end{center}
\caption{\label{inifin} Real (continuous line) and imaginary (dashed
line) parts of $\phi$ in a 1D system of size 1 and periodic boundary
conditions.  Top: the initial condition given by filtered Gaussian
noise. Bottom: final state for $\tilde{t}=5$.  Parameters are
$\varepsilon = 0.0011$, $\alpha = 0.5$ and $\omega=2$.}
\end{figure}

The initial condition for $\phi$ is filtered Gaussian noise of
amplitude 1. Since Eq.\ (\ref{CSHscaled}) does not include spatial
scales smaller than $\delta_{\mathrm{diff}} \sim \sqrt{\varepsilon}$,
the modes with wavenumber greater than $k \sim 1/\sqrt{\varepsilon}$
are initially filtered.  The initial condition for $(E,P,N)$ is
obtained from the one for $\phi$ using Eq.\ (\ref{EPNexpansion}). In
Fig.\ \ref{inifin} we show the real and imaginary part of $\phi$ at
$\tilde{t}=0$ and $\tilde{t}=5$, for $\varepsilon = 0.0011$, $\alpha
= 0.5$ and $\omega=2$.

\begin{figure}[t]
\begin{center}
\includegraphics[width=0.9\columnwidth,keepaspectratio]{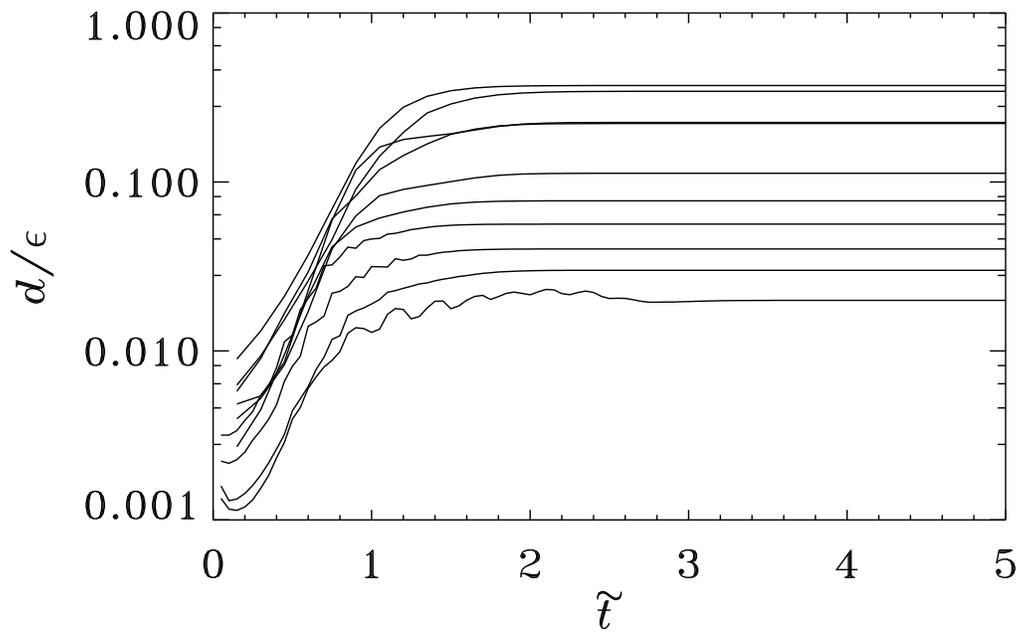}
\end{center}
\caption{\label{da05w2} Average relative error $d/\varepsilon$ in log
scale against time $\tilde{t}$ for different values of $\varepsilon$.
From top to bottom, $\varepsilon = 0.025, 0.018, 0.013, 0.0088,
0.0063, 0.0044, 0.0031, 0.0022, 0.0016$ and $0.0011$. Parameters are
$\alpha = 0.5$ and $\omega=2$}
\end{figure}

In order to calculate an average relative error, we divide $d$ by
$\varepsilon$, the typical magnitude of the fields in the Maxwell
Bloch equations (note that $\phi$ is of order 1).  In Fig.\
\ref{da05w2} we plot $d/\varepsilon$ in log scale against
$\tilde{t}$, for different values of $\varepsilon$ and for $\alpha =
0.5$ and $\omega=2$.

\begin{figure}[t]
\begin{center}
\includegraphics[width=0.9\columnwidth,keepaspectratio]{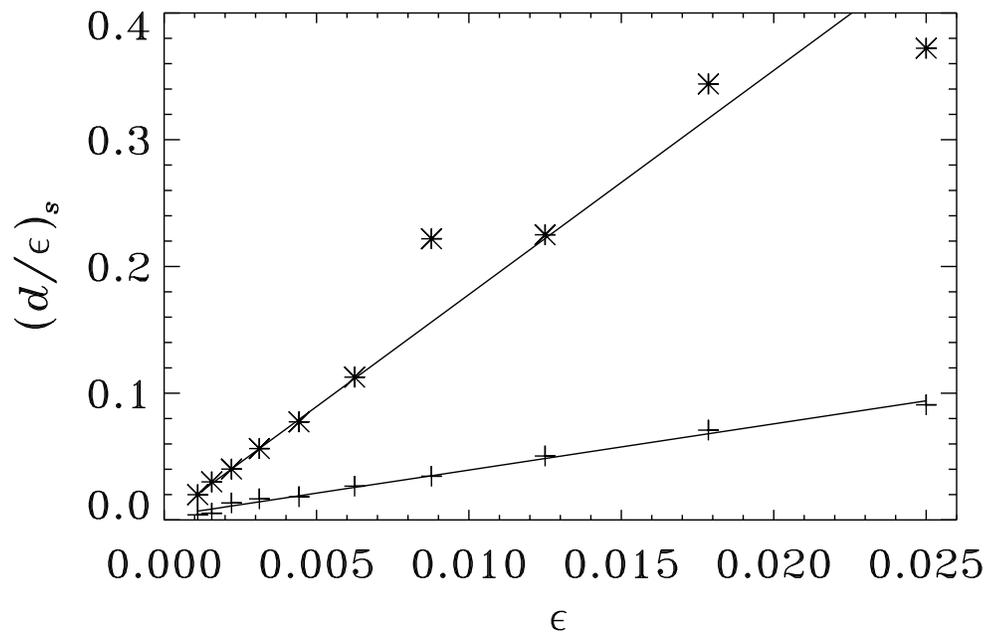}
\end{center}
\caption{\label{dvse} Stationary relative error $(d/\varepsilon)_s$
against $\varepsilon$. Plus symbols correspond to $\alpha = 0.75$,
$\omega = 0.5$ (slope $3.6\pm 0.1$); and asterisks to $\alpha = 0.5$,
$\omega = 2$ (slope $18\pm 1$).}
\end{figure}

We consider the stationary value of the relative error for long times
$(d/\varepsilon)_s$, and check the theoretical asymptotic behaviour
$(d/\varepsilon)_s \sim \varepsilon$.  In Fig.\ \ref{dvse} we plot
$(d/\varepsilon)_s$ against $\varepsilon$ for two points in parameter
space: $\alpha = 0.75$, $\omega = 0.5$; and $\alpha = 0.5$, $\omega =
2$.  In both cases, the linear behaviour $(d/\varepsilon)_s \sim
\varepsilon$ is confirmed.  But the numerical result offers a new
important figure: the slope.  The slopes are $3.6\pm 0.1$ for $\alpha
= 0.75$, $\omega = 0.5$; and $18 \pm 1$ for $\alpha = 0.5$, $\omega =
2$. The increase of the slope in the second case is related to the
fact that patterns with smaller length scales appear for $\alpha =
0.5$, $\omega = 2$ (see Fig.\ \ref{apos}).

An experimental confirmation of the CSH equation would require to
know a specific value of the appropriate distance to threshold for
which the equation is valid.  Let us suppose that the sought
experimental confirmation has a maximum relative error of 10\% and
make the favorable assumptions that the Maxwell Bloch equations
accurately describe the experiment and that the chosen parameters
correspond to a simple pattern with characteristic length equal to
$\delta_{\mathrm{disp}} \sim 1$ as, for example, for $\alpha = 0.75$,
$\omega = 0.5$.  Then, using the slopes of Fig.\ \ref{dvse}, we can
calculate that the distance to threshold should not exceed $r-1 =
0.003$, and the pump must be tuned with a relative error smaller than
0.3\%. For $\alpha = 0.5$, $\omega = 2$, where patterns with
diffusive scale $\delta_{\mathrm{diff}} \sim \sqrt{\varepsilon}$
arise, the situation is worse since the maximum distance to threshold
is $r-1 = 0.0006$, and the relative error of the pump should be
smaller than 0.06\%.

\section{Conclusions}

We performed numerical integrations of the Maxwell Bloch equations
and the corresponding CSH equation, for a class C laser.  The CSH
equation gives a simplified and reduced dynamics of the original
Maxwell Bloch equations for small detuning.  Comparing the results
produced by both set of equations, we obtain an average relative
error of the CSH equation solutions.  The numerical results confirm
the following theoretical prediction: $(d/\varepsilon)_s \sim
\varepsilon$, where $(d/\varepsilon)_s$ is the stationary relative
error that is reached for long times (the small parameter
$\varepsilon$ is introduced in the deduction of the CSH equation and
is directly related to the detuning and distance to threshold).
Therefore, as expected, the CSH equation with terms of different
asymptotic order \cite{martelhoyuelos06} is the appropriate envelope
equation to accurately represent the behaviour of the Maxwell Bloch
equations when $\varepsilon \rightarrow 0$.

The numerical results also allow us to estimate the distance to
threshold range for which the CSH equation holds.  Assuming an
average relative error of 10\%, the maximum distance to threshold is
between 0.003 and 0.0006 for the parameter values analyzed: $\alpha =
0.75$, $\omega = 0.5$, and $\alpha = 0.5$, $\omega = 2$.  The most
restrictive value (0.0006) corresponds to the case when the resulting
pattern has diffusive scales ($\delta_{\mathrm{diff}} \sim
\sqrt{\varepsilon}$).  Although it is not unfeasable to
experimentally establish such small distance to threshold, it
requires a fine tuning of the pump that is not usually available in
standard lasers. Another important experimental difficulty is to
obtain a wide enough beam for the patterns to develop.

Finally, it is important to mention that, despite of the problems for
setting up an accurate laser experiment in the CSH range,  the
numerical results on this paper confirming the validity of the CSH
equation are interesting and valuable from the more general point of
view of Pattern Formation. The CSH equation is an envelope equation
and it is universal in the sense that its structure depends only on
the kind of instability of the problem and not on the particular
physical problem under consideration.

\begin{acknowledgments}
This work has been supported by AECI (Agencia Espa\~nola de
Cooperaci\'on Internacional), Spain, under grant PCI A/6031/06. M.H.
acknowledges Consejo Nacional de Investigaciones Cient\'{\i}ficas y
T\'{e}cnicas (CONICET, PIP 5666, Argentina) and ANPCyT (PICT 2004, N
17-20075, Argentina) for partial support. The work of C.M. has been
supported by the Spanish Ministerio de Educaci\'on y Ciencia under
grant MTM2007-62428, by the Universidad Polit\'ecnica de Madrid under
grant CCG07-UPM/000-3177, and by the European Office of Aerospace
Research and Development under grant FA8655-05-1-3040.
\end{acknowledgments}

\bibliographystyle{apsrev}

\newpage

\end{document}